\documentclass[pra,aps,twocolumn,psfig,epsfig,showpacs,showkeys,floatfix,superscriptaddress]{revtex4-1}
\usepackage{color}

\RequirePackage{amsfonts}

\usepackage{colordvi}

\usepackage{graphicx,amssymb,epsfig,amsmath}
\usepackage{epstopdf}

\newcommand{\be}{\begin{equation}}
\newcommand{\ee}{\end{equation}}
\newcommand{\bea}{\begin{eqnarray}}
\newcommand{\eea}{\end{eqnarray}}

\begin{document}
\title{Quench dynamics of 1D Bose gas in an optical lattice: does the system relax?}

\author{S. Bera}
\affiliation{Department of Physics, Presidency University, 86/1 College Street,
Kolkata 700083, India}

\author{R. Roy}
\affiliation{Department of Physics, Presidency University, 86/1   College Street,
Kolkata 700083, India}

\author{A. Gammal}
\affiliation{Instituto de F\'isica, Universidade de S\~{a}o Paulo, CEP 05508-090, S\~{a}o Paulo, Brazil }

\author{B. Chakrabarti}
\affiliation{Department of Physics, Presidency University, 86/1 College Street,
Kolkata 700083, India}
\affiliation{Instituto de F\'isica, Universidade de S\~{a}o Paulo, CEP 05508-090, S\~{a}o Paulo, Brazil }

\author{B. Chatterjee}
\affiliation{Department of Physics, Indian Institute of Technology-Kanpur, Kanpur 208016, India}
\date{\today}

\begin{abstract}
Understanding the relaxation process is the most important unsolved problem in non-equilibrium quantum physics. Current understanding primarily concerns on if and how an isolated quantum many-body system thermalize. However, there is no clear understanding of what conditions and on which time-scale do thermalization occurs. In this article, we simulate the quench dynamics of one-dimensional Bose gas in an optical lattice from an {\it {ab initio}} perspective by solving the time-dependent many-boson Schr\"odinger equation using the  multi-configurational time-dependent Hartree method for bosons (MCTDHB). We direct a superfluid (SF) to Mott-insulator (MI) transition by performing two independent quenches: an interaction quench when  the interaction strength is changed instantaneously, and a lattice depth quench where the depth of the lattice is altered suddenly. We show that although the Bose-Hubbard model predicts identical physics, the general many-body treatment shows significant differences between the two cases. We observe that lattice depth quench exhibits a large time-scale to reach the MI state and shows an oscillatory phase collapse-revival dynamics and a complete absence of thermalization that reveals through the analysis of the time-evolution of the reduced one-body density matrix, two-body density, and entropy production. In contrast, the interaction quench shows a swift transition to the MI state and shows a clear signature of thermalization for strong quench values. We provide a physical explanation for these differences and prescribe an analytical fitting formula for the time required for thermalization.

\end{abstract}

\maketitle

\section{Introduction} \label{intro}
The relaxation of an isolated quantum many-body system is one of the most important and challenging problems in the field of   non-equilibrium quantum dynamics~\cite{Co:1,Ma:2,Mi:3,Ry:4}. While the maximum entropy principle suggests which types of quantum systems should approach equilibrium, the mechanism of how the system dynamically equilibrate is not known in much detail~\cite{Et:5}. Experimentally, the controlled monitoring and analysis of the statistical relaxation process require the set-up of precise and well-controlled isolated systems. 
Ultracold atoms in a trap is an ideal platform for such experiments as they provide true isolation from the environment as well as enable control of most experimental parameters with arbitrary precision. One-dimensional (1D) system is especially important to study the quantum many-body effect out of equilibrium since the quantum effects are typically more prominent compared to three-dimensional systems~\cite{RMP83:6}. Moreover, from an experimental point of view, 1D systems additionally offers the most precise control of the system parameters~\cite{Langen:7,nat453:8,science:9,nat440:14}.\\

Experiments in non-equilibrium dynamics with ultracold gases reveal a variety of fascinating effects.   
In the landmark experiment by Greiner {\it et al.}~\cite{nat419:11,nat465:12}, the quantum phase transition from superfluid (SF) to Mott-insulator (MI) phase was achieved. There, the atoms prepared in an optical lattice was brought out of equilibrium by quenching the lattice depth and the long-lived coherent collapse and revival dynamics were observed.

Trotzhy {\it et al.}~\cite{natphys:13} demonstrated the rapid relaxation of quasi-local densities, currents, and coherence in strongly correlated 1D Bose gas in an optical lattice. On the other hand Kinoshita {\it et al.}~\cite{nat440:14} showed that the long-lived oscillations in the momentum space implied the absence of thermalization in 1D $^{87}Rb$ atoms.

While experiments on the non-equilibrium dynamics with ultracold gases uncover the question of relaxation in a variety of systems, the microscopic mechanism are still unclear. Importantly, even the understanding of relaxation time-scale and its relation to the system parameter is not understood. Related to the above, the question of entropy production and the effect of the many-body correlations remain an open problem. 
In this work, we explore the relaxation dynamics from a microscopic quantum many-body perspective and answer the fundamental questions on how entropy production is achieved and how the thermalization process relates to the coherence property of the system. We simulate the relaxation dynamics of interacting bosons in 1D optical lattice by solving the time-dependent  Schr\"odinger equation using the {\it{ multi-configurational time-dependent Hartree method for bosons }} (MCTDHB)~\cite{mctdhb:14a,mctdhb:14b}. \\
Bosons in a lattice are generally studied using the Bose-Hubbard model (BHM)~\cite{bhm:15a,bhm:15b}. In the BHM, the quantum phases - the superfluid (SF) and Mott-insulator (MI) are determined by the relative strength of the tunneling coupling $J$ and the interaction parameter $U$. Thus the phase transition is determined by the single parameter ($\frac{U}{J}$)~\cite{bhm:15a}. The BHM, however, does not differentiate between the experimental process of tuning the $U,J$ parameter. Thus the $SF \rightarrow MI$ transition would be achieved for a high $\frac{U}{J}$ value irrespective of how the ratio is achieved.
In experiments, one can tune the ratio by changing either the lattice depth or the interaction strength, and the BHM predicts identical physics. \\
In this work, we simulate the $SF \rightarrow MI$ transition using two independent quenches: (i) quenching (i.e. instantaneous increase) the interactions keeping lattice depth constant and (ii) quenching the lattice depth keeping interactions constant.
We show that although the BHM predicts identical physics, the {\it{ab-initio}} quantum many-body simulations show a significant difference in the quench dynamics between the two set-ups. We initially report the quench dynamics for both lattice depth quench and interaction quench for the same excitation energy and characterize how they are fundamentally   different. In general, we discover that for interaction quench the system exhibits a clear signature of thermalization. However, for the lattice-depth quench, with same excitation energy, we find no signature of thermalization even for long-time dynamics. To conclude whether the system at  all  thermalize  or  not for lattice depth quench, we perform a very strong quench and do not find any clear signature of thermalization in the dynamics. We define two fundamental time-scales of the system: characteristic time $t_{c}$ (the time required to reach the MI phase) and the equilibration time $t_{equi}$ ( the time needed to thermalize). We analyze the reduced one-body density matrix and the two-body density and demonstrate that the coherence properties show fundamentally   different picture between the two set-ups that also reveal through the evolution of both the natural occupations and Shannon entropy. 

We observe that the Shannon entropy $S(t)$ exhibits collapse-revival dynamics even for the strong lattice depth quench when the system is supplied with very  high excitation energy. Entropy dynamics also oscillates for very small interaction quench. However, for stronger interaction quench, $S(t)$ shows a linear increase at a small time-scale followed by clear saturation which assures the thermalization process. We observe that for stronger interaction quench, $t_{equi}$ is simply   $t_{c}$ and $t_{equi}$ follows a power-law decay with an increase in the strength of the interaction quench. \\
The paper is organized as follows. In Section~\ref{method}, we give a brief introduction to the numerical many-body method MCTDHB. In Section~\ref{key_quantities},  we introduce   the key measures in the dynamics that are analyzed subsequently. In Section~\ref{quench_result}, we present our numerical results and in Section~\ref{conclusion}, we conclude our work.   

\begin{figure}[b]
	\begin{center}
		\includegraphics[height=.50\textwidth,angle=-90]{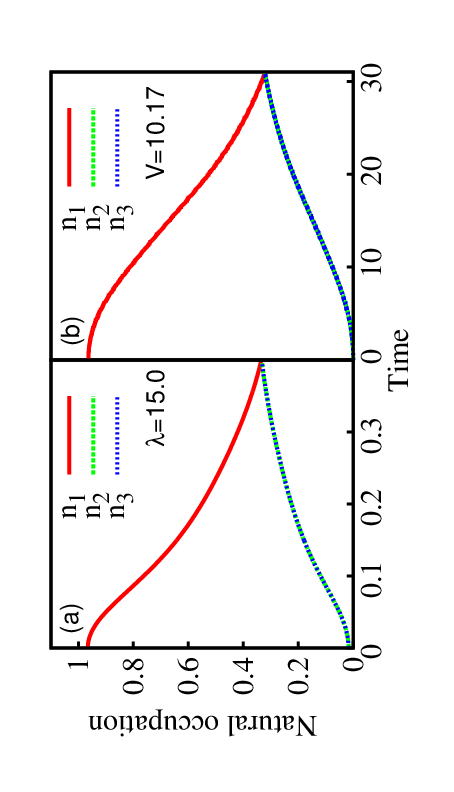}
	\end{center}
	\caption{[panel (a)] Natural-orbital occupations for interaction quench to $\lambda=15.0$. The initial condensed SF phase fragments with time and all three natural orbitals are occupied. At time $t$=0.4, the system becomes fully fragmented MI phase with 33.33 \% natural occupation in three orbitals. [panel (b)] Natural-orbital occupations for lattice-depth quench to $V=10.17$. The initial SF phase fragments with time and all three natural orbitals are  occupied. At time $t$=31.0, the system becomes fully fragmented MI phase with 33.33 \% natural occupation in three orbitals. All quantities are dimensionless. }       
	\label{Fig1}
\end{figure}

\section{Numerical method} \label{method}
We solve the time-dependent Schr\"odinger equation for $N$-interacting bosons: 
\begin{equation}
\hat{H} \Psi = i \frac{\partial \Psi}{\partial t}
\label{TDSE} 
\end{equation}
using the MCTDHB method~\cite{mctdhb:14a,mctdhb:14b} as implemented in the MCTDH-X package~\cite{axel1,axel2,ultracold:20}. Here, the many-body wave function expanded as the linear combination of time-dependent permanents with the time-dependent weights:   
\begin{equation}
\vert \psi(t) \rangle = \sum_{\vec{n}}C_{\vec{n}}(t) \vert \vec{n};t \rangle
\end{equation}
 where, in the second quantized representation 
\begin{equation}
\qquad \vert \vec{n};t\rangle = \prod_{i=1}^M \left[\frac{\left(
	\hat{b}_i^\dagger(t)\right)^{n_i}}{\sqrt{n_i!}} \right] \vert vac \rangle.
\label{second_quan}
\end{equation}
The summation run over all-possible configurations $N_{conf}=\binom{N+M-1}{N}$. Note  that the expansion coefficients $\lbrace C_{\vec{n}} (t); \sum_i n_i = N
\rbrace$ \textit{and} the orbitals $\lbrace \phi_i(x,t)
\rbrace_{i=1}^M$ that build up the permanents are time dependent and fully variationally optimized quantities. To determine the wave function $\vert \psi(t) \rangle$, we need to determine the time evolution of  both the coefficients and the orbitals. Requiring the stationarity of the action functional with respect to variations of the time-dependent coefficients and the set of time-dependent orbitals, we derive their equations of motion. The coupled set of non-linear integro-differential equations are solved simultaneously  to obtain the wavefunction. 
The MCTDHB method has been established as one of the most efficient algorithm to solve the time-dependent many-body problem of interacting bosons accurately and for wide range of problems~\cite{sakmannthesis:16,Axel:Thesis:17,jchem:18,pra88:19,chatterjee15,chatterjee17,chatterjee18,rhom97,bera18}. The efficiency of this method results from the variationally optimized time adaptive basis which makes the sampled Hilbert space dynamically follow the motion of the many-body dynamics. 
Note that in the limit of $M\rightarrow \infty$ the set of permanents $\{\vert \vec{n};t \rangle\}$ span the whole Hilbert space and the expansion is exact. However for practical calculation, we require to limit the size of the Hilbert space to the suitable level of required accuracy. 
\begin{figure}
	\begin{center}
		\includegraphics[height=.50\textwidth]{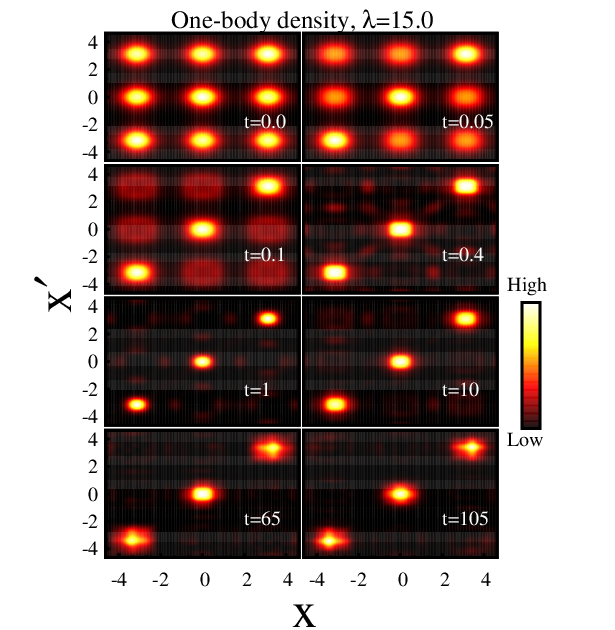}
	\end{center}
	\caption{Time evolution of the reduced one-body density matrix $\vert{\rho^{(1)}(x^{\prime},x)}{\vert}^{2}$ for interaction quench $\lambda=15.0$. We observe very fast relaxation to MI phase and the system thermalizes. Details are in the text. All the quantities are dimensionless.}
	\label{Fig2}
\end{figure}

\begin{figure}
	\begin{center}
		\includegraphics[height=.50\textwidth]{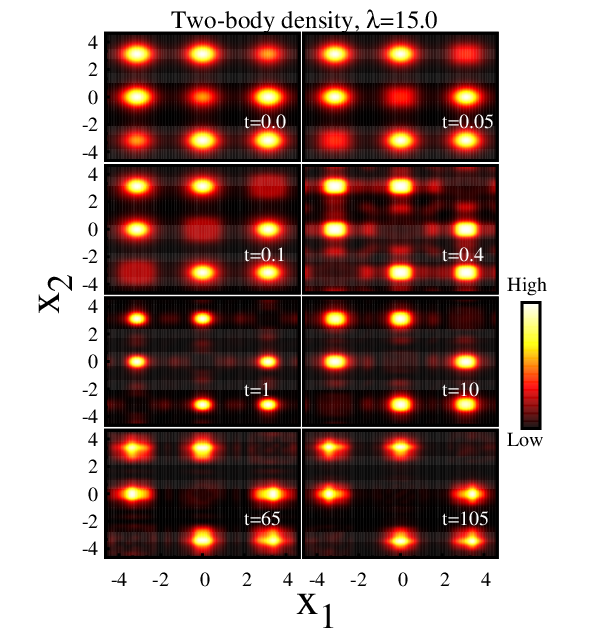}
	\end{center}
	\caption{Time evolution of the two-body density $\rho^{(2)}(x_{1},x_{2})$ for interaction quench $\lambda=15.0$. Details are in the text. All the quantities are dimensionless.}
	\label{Fig3}
\end{figure}

\section{Quantities of interest}\label{key_quantities}
In this section, we define the quantities that we analyze in our subsequent simulations.

From the many-body wave function $\psi(x_{1},x_{2},...x_{N};t)$, the reduced one-body density matrix (RDM) $\rho^{(1)}$ is calculated as
\begin{equation}
\begin{split}
\rho^{(1)}(x^{\prime}\vert x;t)=N\int_{}^{}dx_{2}dx_{3}...dx_{N}\psi^{*}(x^{\prime},x_{2},\dots,x_{N};t) \\ \psi^{*}(x,x_{2},\dots,x_{N};t)
\end{split}
\end{equation}

Its diagonal gives the one-body density $\rho(x,t)$  obtained as 
\begin{equation}
\begin{split}
\rho(x,t)=\int_{}^{}dx_{2}dx_{3}...dx_{N}\psi^{*}(x,x_{2},\dots,x_{N};t) \\ \psi(x,x_{2},\dots,x_{N};t)
\end{split}
\end{equation}
and gives the probability of a particle at the position $x$ at time $t$ when the contribution of other particles are traced out.

The non-local correlations are determined by the off-diagonal kernel of $\rho^{(1)}(x^{\prime}\vert x;t)$. When $\vert x-x^{\prime} \vert \rightarrow \infty$, the off-diagonal behavior of $\rho^{(1)}(x^{\prime},x)$ measures the coherence. Infinite homogeneous system exhibits non-vanishing off-diagonal long-range order (ODLRO). However, for our finite system true ODLRO does not exist and coherence is determined by the behavior of the off-diagonal $\rho^{(1)}(x^{\prime},x)$. \\
The second order reduced density matrix is defined as
\begin{equation}
\begin{split}
\rho^{(2)}(x_{1}^{\prime}, x_{2}^{\prime} \vert x_{1},x_{2};t)=N(N-1)\int_{}^{}dx_{3}dx_{4}....dx_{N}\\ \psi^{*}(x_{1}^{\prime},x_{2}^{\prime},x_{3},...,x_{N};t) \psi(x_{1},x_{2},...x_{N};t)
\end{split}
\end{equation} 
Its diagonal kernel is given by
\begin{equation}
\rho^{(2)}(x_{1},x_{2};t) \equiv \rho^{(2)}(x_{1}^{\prime}=x_{1},x_{2}^{\prime}=x_{2} \vert x_{1},x_{2};t)
\end{equation}
and represents the simultaneous probability of finding a boson at $x_1$ and another at $x_2$. \\
Diagonalizing the one-body RDM, one obtains the eigenvalues known as the natural occupations $n_i$ and eigenvectors called the natural orbitals $\varphi_i$. The many-body Shannon information entropy can be defined in terms of the (time-dependent) natural occupation~\cite{rhom97}
\begin{equation}
S(t)=S^{occu}(t) = -\sum_{i} {\bar{n}}_{i}(t) \Big[ \ln \hspace*{.1cm} {\bar{n}}_{i}(t) \Big] .
\label{Socc}
\end{equation}
where $\bar n_{i}(t)=\frac{n_{i}}{N}.$ 
For Gross-Pitaevskii mean-field theory $S^{occu}(t)=0$ as only one natural occupation contributes, $\bar n_{1}=\frac{n_{1}}{N}=1$.

\begin{figure}  
	\begin{center}
		\includegraphics[height=.50\textwidth,angle=-90]{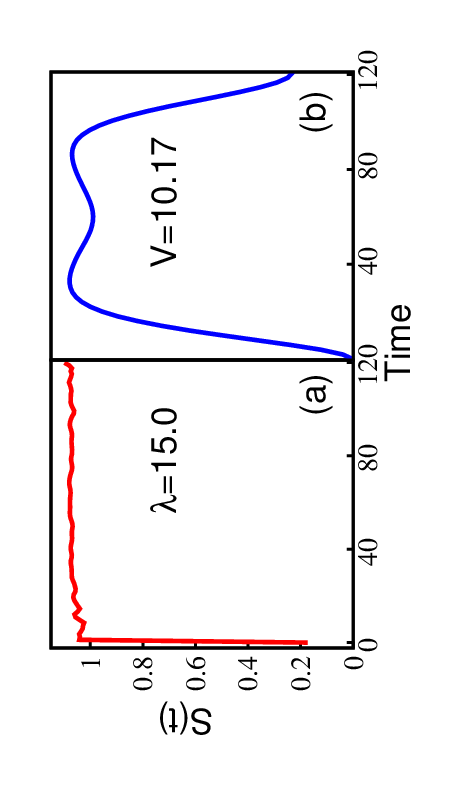}
	\end{center}
	\caption{Panel(a) Dynamics of Shannon entropy $S(t)$ for interaction quench $\lambda=15.0$ The sharp linear increase followed by saturation signify the thermalization process. The saturation value is $=1.098$. Panel(b) Dynamics of Shannon entropy $S(t)$ for lattice-depth quench $V=10.17$. The oscillations in the entropy production rules out the possibility of thermalization. All the quantities are dimensionless.}
	\label{Fig4}
\end{figure}

\begin{figure}
	\begin{center}
		\includegraphics [height=.4\textwidth, angle=-90]{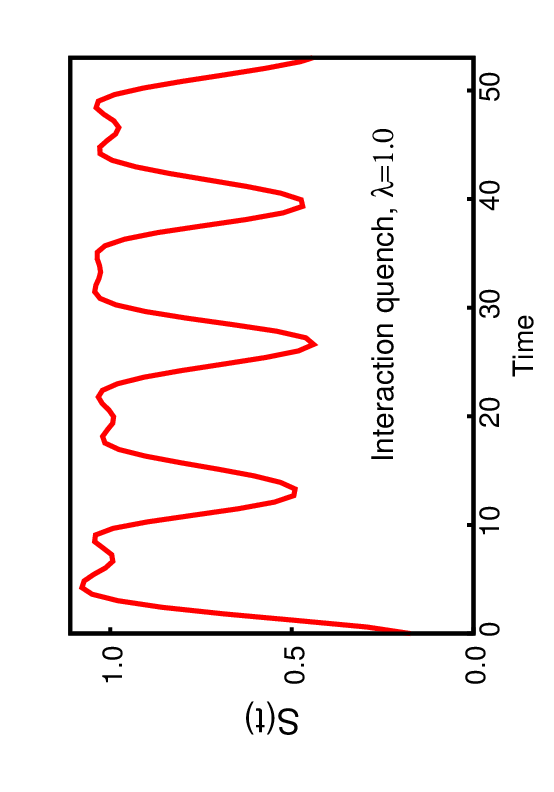}
	\end{center}
	\caption{ Dynamics of Shannon entropy $S(t)$ for interaction quench $\lambda=1.0$. The oscillations in the entropy production rules out the possibility of thermalization. All the quantities are dimensionless.}
	\label{Fig5}
\end{figure}

\section{Results for quench dynamics}\label{quench_result}

Our setup consists of $N=3$ bosons in a one-dimensional triple well optical lattice $S=3$ with periodic boundary condition modeled as $V_{OL}=V \sin^{2}(kx)$, $k$ is the wave vector, and $V$ is the depth of the lattice. The bosons interact with a contact interaction $\hat{W}(x_{i}-x_{j})=\lambda \delta(x_{i}-x_{j})$ where $\lambda$ is the dimensionless strength. Experimentally the optical lattice is constructed using a pair of counter propagating laser beams that creates a periodic standing wave pattern. The depth of the lattice is experimentally changed by varying the laser intensity while the interaction strength is tuned using Feshbach resonances~\cite{feshbach:22a}. This simple set-up captures all the necessary physics required for the current problem and can be generalized for bigger systems.  

This triple-well set-up can be addressed using a three-site Bose-Hubbard model

\begin{equation}
\hat {H}_{BH} = -J\sum_{i,j}^{}{ \hat {b_{i}}^{\dagger}b_{j}} + \frac{U}{2}\sum_{i}^{}{\hat{ n}_{i}(\hat {n_{i}} -1)}
\end{equation}

where $U$ is the on-site interaction energy and $J$ is the hopping matrix. The competition between the two terms $J$ and $U$ give rise to the emergent phases -- the superfluid(SF) and the Mott insulator(MI) phase. Crucially the phase transition $SF\rightarrow MI$ is determined solely by the ratio of $J$ and $U$. For a homogeneous square lattice the critical ratio for the transition is $\frac{U}{J} = 0.3$~\cite{prb53:23,prb61:24}. However, from a general many-body perspective, the simple reduction of the physics to a single parameter becomes inaccurate. We now perform two separate quench - an interaction quench and a lattice depth quench and demonstrate the difference between the two set-ups.

\begin{figure}
	\begin{center}
		\includegraphics [height=.5\textwidth, angle=0]{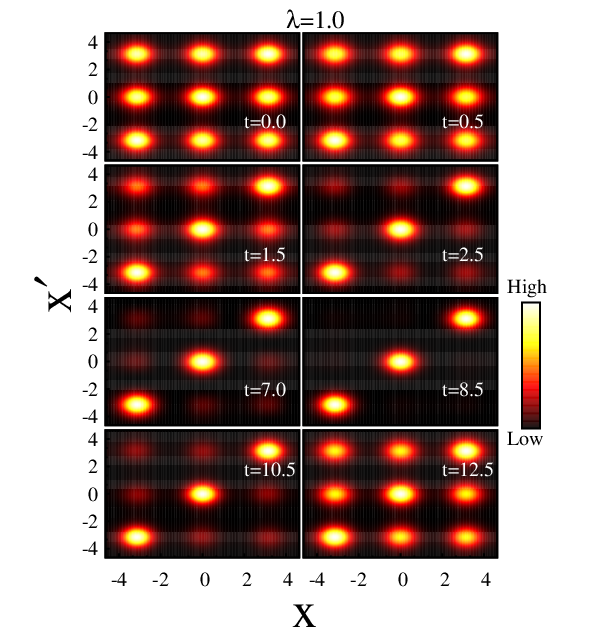}
	\end{center}
	\caption{Time evolution of the reduced one-body density matrix $\vert{\rho^{(1)}(x^{\prime},x)}{\vert}^{2}$ for interaction quench $\lambda=1.0$. Collapses and revivals demonstrate how the matter-wave field dephases and rephases periodically during its time evolution. The system does not thermalizes. All the quantities are dimensionless.}
	\label{Fig6}
\end{figure}

\begin{figure}
	\begin{center}
		\includegraphics [height=.5\textwidth, angle=270]{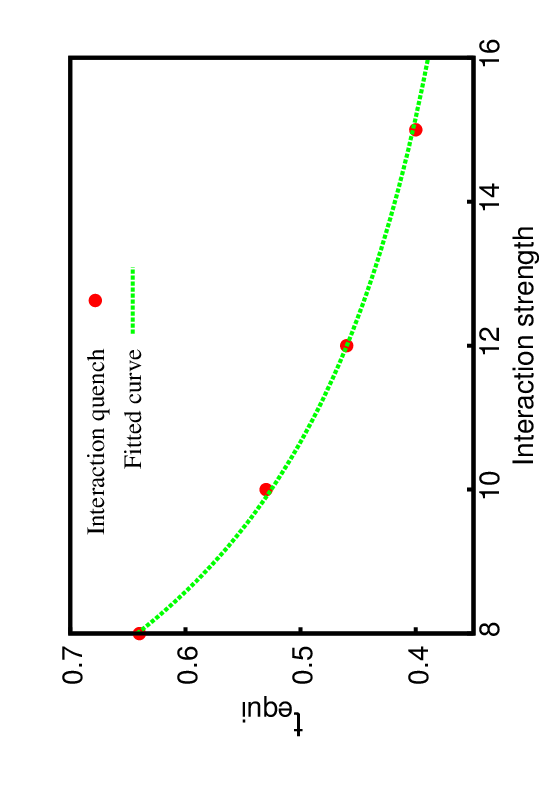}
	\end{center}
	\caption{Equilibration time $t_{equi}$ for interaction quench. The curve shows power-law decay. The best fit formula is $t_{equi} =4.498\lambda^{-0.9301}$. All the quantities are dimensionless.}
	\label{Fig7}
\end{figure}

\subsection{Interaction quench}\label{interaction_quench}
For the interaction quench, the interaction strength $\lambda$ is changed suddenly keeping the depth of the lattice $V$ fixed. Our main motivation is to contrast the system response between the lattice depth quench and interaction quench. To highlight the difference with respect to the BHM, we perform both quenches the for same excitation energy. In both quenches, the system makes a dynamical transition from $SF \rightarrow MI$ phase but with a different time-scale. To achieve the $SF\rightarrow MI$ transition with interaction quench, the ground state is constructed with depth $V=3.0$ and $\lambda=0.1$.The SF phase is characterize by global correlation across the lattice sites. Keeping $V$ fixed, $\lambda$ is changed instantaneously to $\lambda=15.0$. This corresponds to the pumping the system with energy $E=5.0$ through two-body interaction term in the Hamiltonian. \\ In Fig.~\ref{Fig1}(a), we plot natural orbital occupation as a function of time. Initially, $(t=0)$ only the first natural orbital contributes, which corresponds to the SF phase. Here the many-body wave-function is equivalent to the mean-field wave-function which can be represented as $\vert {3,0,0} \rangle$. With an increase in time, fragmentation starts, and at $t=0.4$ the system shows completely  three-fold fragmented state with $33.33 \%$ population in each orbital. This three-fold fragmented state corresponds to the MI phase and is described as $\vert 1,1,1 \rangle$. We define characteristic time $t_{c}$ as the time required by the system to make a transition from the condensed SF state to the fragmented MI state. For the present set-up the characteristic time $t_{c}=0.4$. \\
Fig.~\ref{Fig2} presents the reduced one-body density matrix for different times. Initially, at $t=0$, we observe a uniform distribution of maximas. This shows that the initial state (which is SF) displays both intra-well as well as inter-well phase coherence. As time $t$ increases, the off-diagonal maxima fade out and the diagonal contributions become more pronounced. At the characteristic time $t_{c}=0.4$,  which corresponds to the equivalent MI phase, only the diagonal maxima are observed with a complete absence of off-diagonal contributions showing a complete absence of phase-coherence. From long-time dynamics, we do not observe any revival of coherence and can conclude that the system thermalizes. The corresponding two-body density $\rho^{(2)}(x_{1},x_{2})$ are shown in Fig.~\ref{Fig3}. At $t=0.05$, the diagonal maximas shows a reduction in amplitude compared to the off-diagonal maxima. However at time $t=0.4$ an equal distribution per site is achieved although the diagonal does not extinguish completely. With time, the diagonal contributions fade and complete depletion of the diagonal is achieved at a much larger time ($t=105$). The corresponding Shannon information entropy as a function of time is shown in Fig.~\ref{Fig4}(a), We observe a generic linear increase at a shorter time followed by saturation. The sharp linear increase in $S(t)$ is attributed to an exponential increase in the time-dependent natural occupation contributing to the dynamics. This clearly implies the onset of chaos and thermalization~\cite{therm1,therm2}. The saturation of $S(t)$ happens due to the complete occupation of the available finite sized Hilbert space.
For Gaussian orthogonal ensemble (GOE) of random matrices the saturation value of information entropy is given by $S_{GOE} = -\sum_{i=1}^{M} (\frac{1}{M})\ln (\frac{1}{M})=\ln(M)=1.099$ where $M$ is the number of orbitals~\cite{kota:rmt,mehta:rmt,haake:rmt}. The saturation value of Shannon entropy for our present simulation is $S^{sat}= 1.098$. This value of saturation closely matches with that of the GOE predicted value. Since the GOE entropy saturation implies thermalization we can clearly infer that here the system thermalizes. We define the equilibration time $t_{equi}$ as the time required by the system to thermalize. Here $t_{equi} = 0.4$. \\ 

In Table~\ref{tab:title1}, we report characteristic time $t_{c}$ and the equilibration time $t_{equi}$ for different values of $\lambda$. Both $t_{c}$ and $t_{equi}$ decreases as the interaction quench strength $\lambda$ increases. Interestingly, for small $\lambda \leqslant 6.0$ quench, there is an absence of thermalization. To demonstrate, we show the temporal evolution of $S(t)$ for a quench with $\lambda=1$ in Fig.~\ref{Fig5}. We observe that instead of saturation, the information entropy $S(t)$ shows a periodic oscillatory behavior thus showing an absence of thermalization. The corresponding dynamics of the reduced one-body density matrix is displayed in Fig.~\ref{Fig6}. Unlike that of $\lambda=15.0$ quench, here we see a periodic collapse-revival cycle with time and the complete diagonal depletion is seen before is not observed and thus the possibility of thermalization is ruled out. Here, since $\lambda$ is small, the system receives only a small amount of energy and the interaction quench essentially  acts as an external perturbation. 
Thus from Table~\ref{tab:title1}, we observe that thermalization is possible only for stronger interaction and the corresponding equilibration time $t_{equi}$ decreases with an increase in $\lambda$. In Fig.~\ref{Fig7}, we plot $t_{equi}$ for different $\lambda$ quenches  and  we have deduced the best fitting formula for $t_{equi}$ as $t_{equi} =4.498\lambda^{-0.9301}$ which shows a power-law decay.\\ The fitting formula from Fig.~\ref{Fig7}, can be rewritten as $t_{equi}\lambda^{0.93} \approx 4.5$. Now we can make an analogy with the classical damped harmonic oscillator. For small $\lambda$, we have {\it {subcritical}} behavior: shows oscillations manifested as collapses and revivals. At {\it {critical}} interaction strength $\lambda_{c} \approx 8.0$ the system starts to thermalize {\it {i.e.}} stops oscillation. For $\lambda>\lambda_{c}$, we have {\it {supercritical}} behavior {\it {i.e,}} clear signature of thermalization. Thus comparing with the case of classical damped oscillator, we may conclude that $\lambda$ plays a role of damping parameter in interaction quench. 
\begin{figure} 
	\begin{center}
		\includegraphics[height=.5\textwidth]{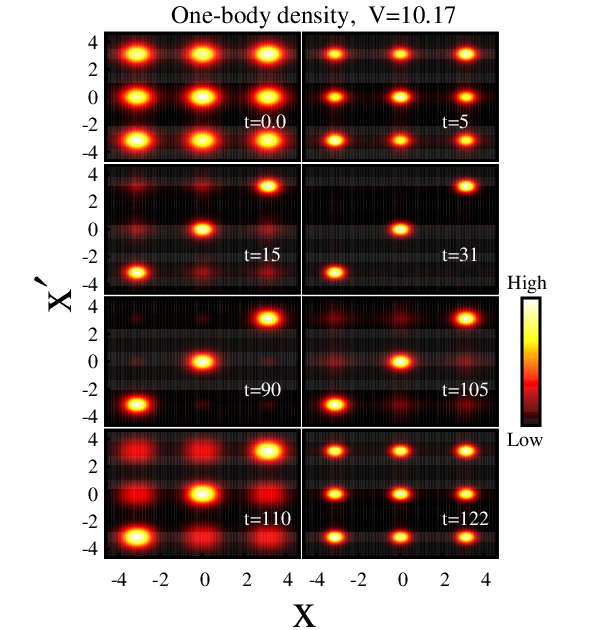}
	\end{center}
	\caption{Time evolution of the reduced one-body density matrix $\vert{\rho^{(1)}(x^{\prime},x)}{\vert}^{2}$ for lattice-depth quench $V=10.17$. Collapses and revivals demonstrate how the matter-wave field dephases and rephases periodically during its time evolution. The system does not thermalizes. All the quantities are dimensionless.}
	\label{Fig8}
\end{figure}
\begin{figure} 
	\begin{center}
		\includegraphics[height=.5\textwidth]{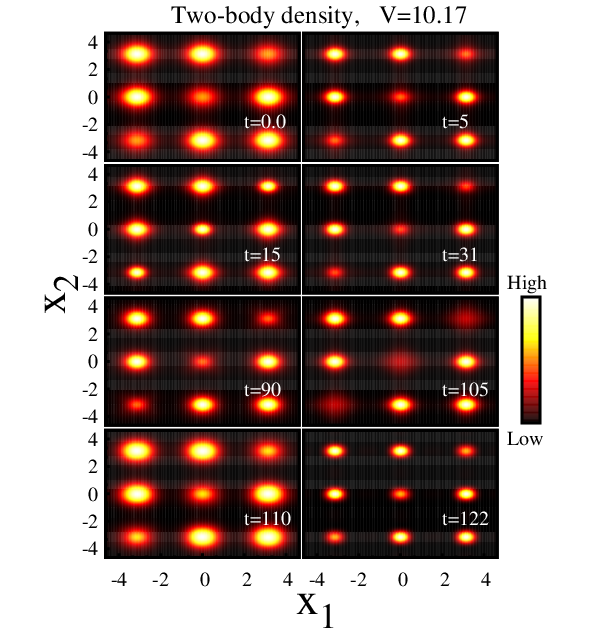}
	\end{center}
	\caption{Time evolution of the two-body density $\rho^{(2)}(x_{1},x_{2})$ for lattice-depth quench $V=10.17$. Details are in the text. All the quantities are dimensionless.}
	\label{Fig9}
\end{figure}
\begin{table}[!]
	\caption {Characteristic time $t_c$, thermalization or equilibration time $t_{equi}$ for different interaction quench $\lambda$. $E$ corresponds to the total energy and $\Delta E$ is the excitation energy} \label{tab:title1} 
	\begin{center}
		\begin{tabular}{|c |c |c |c |c|} 
			\hline
			$\lambda$ & $E$ & $\Delta E$ & $t_{c}$& $t_{equi}$ \\ [0.5ex] 
			\hline
			1.0   & 3.43 & 0.3 & 4.5& No thermalization  \\ 
			\hline
			5.0  & 4.77 & 1.64 & 1.0 & No thermalization\\
			\hline
			6.0 & 5.12  & 1.89 & 0.83  & No thermalization   \\
			\hline
			8.0 &5.78  & 2.65  & 0.64 & 0.64 \\
			\hline
			10.0  & 6.45 & 3.32 & 0.53& 0.53 \\
			\hline
			12.0 &7.13 & 4.0 & 0.46 & 0.46\\
			\hline
			15.0 & 8.13 & 5.0 & 0.40&0.40\\
			\hline
		\end{tabular}
	\end{center}
\end{table}
\begin{table}[!]
	\caption {Characteristic time $t_c$ for different lattice depth quench $V$. $E$ corresponds to the total energy and $\Delta E$ is the excitation energy. since we do not observe any thermalization for lattice depth quench, so we are not quoting any equilibration time $t_{equi}$.} \label{tab:title2} 
	
	\begin{center}
		\begin{tabular}{|c |c |c |c |} 
			\hline
			$V$ & $E$ & $\Delta E$ & $t_{c}$ \\ [0.5ex] 
			\hline
			15  & 11.50 & 8.37 & 31 \\
			\hline
			20 & 14.97  & 11.84 & 29  \\
			\hline
			25 & 17.45  & 14.32 & 26.5 \\
			\hline
			30  & 21.94 & 18.81 & 24  \\
			\hline
			40 & 35.88  &  32.75 & 22.5 \\
			\hline
		\end{tabular}
	\end{center}
\end{table}
\subsection{Optical Lattice depth quench}\label{depth_quench}
We now perform a quench of the lattice depth $V$ keeping the interaction strength $\lambda $ fixed. The BHM predicts identical physics as long as the $U/J$ ratio is the same. However, in the following we show that fundamental difference arises for the two different quenches. We initialize the system in the ground state $\vert \psi(t=0) \rangle$ of the Hamiltonian with $\lambda=0.1$ and $V=3.0$ which corresponds to SF phase as reported in the previous section [~\ref{interaction_quench}] and instantaneously increase the lattice depth to $V=10.17$, where the system is pumped with same excitation energy as mentioned for $\lambda=15.0$ interaction quench. The corresponding natural occupation dynamics is shown in Fig.~\ref{Fig1}(b). It has a similar behavior as the interaction quench [Fig.~\ref{Fig1}(a)], the system initially in the $\vert {3,0,0} \rangle$ state becomes fragmented with time and finally reached fragmented MI phase with configuration $\vert {1,1,1} \rangle$ at the characteristic time $t_{c}=31.0$. For the interaction quench with the same excitation energy $t_{c} = 0.4$ as reported in section ~\ref{interaction_quench}. Thus we find that $t_{c}$ for lattice depth quench is significantly larger compared to the $t_{c}$ of the interaction quench.  
The corresponding dynamics of reduced one-body density matrix is presented in Fig.~\ref{Fig8} which clearly  exhibits  the long-time collapse-revival dynamics as observed in the Greiner experiment~\cite{nat419:11}. The time evolution of the two-body density is shown in Fig.~\ref{Fig9}. The diagonal maximas reduces with time but is never completely depleted and revives at long time-scales. The corresponding Shannon entropy, shown in Fig.~\ref{Fig4}(b), also exhibits the periodic oscillation in time. The absence of a generic linear increase and saturation guarantees that the system will not thermalize. In Table~\ref{tab:title2}, we present the results for different values of lattice depth quench. With increase in lattice depth $V$, the characteristic time $t_{c}$ decreases consistently although compared to the interaction quench, here $t_{c}$ does not drop very sharply with increase in $V$. Even for very very strong quench, we observe the collapse-revival dynamics both in the reduced one-body density matrix and two-body density. The corresponding Shannon entropy also shows oscillatory behavior. All these observations are mutually consistent in ruling out the possibility of thermalization.\\

We have so far explicitly shown the dynamics results for unit filling factor i.e., three atoms  in three wells. We have checked the above results for higher filling factors and have found that our conclusions remain valid. 
\section{Conclusion}\label{conclusion}
In this paper, we have studied the quench dynamics of 1D interacting bosons in an optical lattice from a first-principle general quantum many-body perspective using the MCTDHB method. Our motivation is to observe and understand thermalization with quench dynamics beyond the BHM prediction. We observe that the relaxation process for the two different quenches is vastly different from each other. For strong interaction quench, we observe a clear signature of thermalization that displays in the observation of all the key quantities. The equilibration time exhibits power-law decay with the increase in interaction strength. For smaller interactions, we see an absence of thermalization.
In contrast, the lattice depth quench for the same excitation energy exhibits long-time collapse-revival dynamics for all values of lattice depth. The system smoothly reaches the MI phase and the characteristic time is significantly larger than the corresponding interaction quench. We do not observe any possibility of thermalization even when the excitation energy given is very strong. From these observations, we conjecture that lattice depth quench does not lead to thermalization. 
Although we have explored the thermalization properties in the SF to MI regime in details, many open questions and further investigations remain. Understanding the dynamics in fermionization regime for larger integer filling and incommensurate filling set-ups are possible  extensions.       

\begin{acknowledgments}
	S. Bera wants to acknowledge Department of Science and Technology (Government of India) for the financial support through INSPIRE fellowship [2015/IF150245]. R. Roy acknowledges UGC fellowship. B. Chakrabarti acknowledges FAPESP (grant No. 2016/19622-0) . A. Gammal acknowledges FAPESP and CNPq  for financial support. B. Chatterjee acknowledges financial support from the Department of Science and Technology, Government of India under the DST Inspire Faculty fellowship.
\end{acknowledgments}

\end{document}